\documentclass[conference]{IEEEtran} 
\IEEEoverridecommandlockouts
\pdfoutput=1
\usepackage{CJKutf8}
\usepackage{cmap}
\usepackage[T1]{fontenc}
\usepackage{cite} 
\usepackage{amsmath,amssymb,amsfonts}
\usepackage{graphicx} 
\usepackage{textcomp}
\usepackage{epstopdf}
\usepackage{xcolor}
\usepackage{hyperref}
\usepackage{bm}
\usepackage{amsthm}
\usepackage{enumerate}
\usepackage{booktabs} 
\usepackage{diagbox}
\usepackage{supertabular}
\usepackage{subfigure}
\usepackage{caption}
\usepackage{algorithmicx}
\usepackage[linesnumbered,ruled,vlined]{algorithm2e}

\begin{document} 
\begin{CJK}{UTF8}{gbsn}
\pdfoutput=1
\title{Optimizing Information Propagation for Blockchain-empowered Mobile AIGC: A Graph \\ Attention Network Approach}
%
\author{\IEEEauthorblockN{Jiana Liao\IEEEauthorrefmark{1}, Jinbo Wen\IEEEauthorrefmark{2}, Jiawen Kang\IEEEauthorrefmark{1}, Yang Zhang\IEEEauthorrefmark{2}, Jianbo Du\IEEEauthorrefmark{3},  Qihao Li\IEEEauthorrefmark{5}, Weiting  Zhang\IEEEauthorrefmark{4}, Dong Yang\IEEEauthorrefmark{4}}

\IEEEauthorblockA{\IEEEauthorrefmark{1}\textit{Guangdong University of Technology, China}
\IEEEauthorrefmark{2}\textit{Nanjing University of Aeronautics and Astronautics, China}\\
\IEEEauthorrefmark{4}\textit{Beijing Jiaotong University, China
}
\IEEEauthorrefmark{5}\textit{Jilin University, China}
\IEEEauthorrefmark{3}\textit{Xi'an University of Posts and Telecommunications, China}
}

\IEEEcompsocitemizethanks{The work was supported by the National Natural Science Foundation of China (NSFC) under Grants No. 62102099 and No. U22A2054, the Pearl River Talent Recruitment Program under Grant 2021QN02S643, and the Guangzhou Basic Research Program under Grant 2023A04J1699, and also supported by the NSFC under grant No. 62071343, the Foundation of State Key Laboratory of Public Big Data (No. PBD2023-12), and the Collaborative Innovation Center of Novel Software Technology and Industrialization. (Corresponding author: Jiawen Kang, e-mail: kavinkang@gdut.edu.cn).}
}

\maketitle
\begin{abstract}
Artificial Intelligence-Generated Content (AIGC) is a rapidly evolving field that utilizes advanced AI algorithms to generate content. Through integration with mobile edge networks, mobile AIGC networks have gained significant attention, which can provide real-time customized and personalized AIGC services and products. Since blockchains can facilitate decentralized and transparent data management, AIGC products can be securely managed by blockchain to avoid tampering and plagiarization. However, the evolution of blockchain-empowered mobile AIGC is still in its nascent phase, grappling with challenges such as improving information propagation efficiency to enable blockchain-empowered mobile AIGC. In this paper, we design a Graph Attention Network (GAT)-based information propagation optimization framework for blockchain-empowered mobile AIGC. We first innovatively apply age of information as a data-freshness metric to measure information propagation efficiency in public blockchains. Considering that GATs possess the excellent ability to process graph-structured data, we utilize the GAT to obtain the optimal information propagation trajectory. Numerical results demonstrate that the proposed scheme exhibits the most outstanding information propagation efficiency compared with traditional routing mechanisms.
\end{abstract}

\begin{IEEEkeywords}
Mobile AIGC, information propagation, age of information, graph attention network.
\end{IEEEkeywords}

\section{Introduction}
Artificial Intelligence-Generated Content (AIGC) emerges as a pioneering technique\cite{xu2024unleashing}, positioned at the vanguard of technological innovation. It spearheads a transformative wave in the content creation landscape by leveraging sophisticated natural language processing and machine learning technologies. Departing from traditional manual content generation, AIGC signifies a paradigmatic shift in how AI perceives and produces digital content\cite{xu2024unleashing}, spanning across diverse mediums such as text, images, and audio. However, the current centralized AIGC framework encounters challenges like service latency\cite{wen2023freshness}. To address this challenge, the convergence of AIGC with mobile edge networks has given rise to mobile AIGC networks\cite{xu2024unleashing, zhang2023complete}, which can not only mitigate service latency issues but also ensure that users benefit from more responsive and tailored content suggestions.

With the incredible ability to empower secure and transparent data transactions through decentralized ledgers, blockchains have attracted significant attention from both academia and industry\cite{kang2023blockchain}. Based on consensus mechanisms and encryption technologies, blockchains play a pivotal role across various domains, such as metaverses and federated learning\cite{kang2023blockchain}. In mobile AIGC networks with multiple AIGC service providers and multiple AIGC users, blockchain technologies are designed to ensure the data secure storage and efficient transactions of AIGC products. Therefore, blockchains are considered fundamental and indispensable technologies for the advancement of mobile AIGC.

Although blockchain-empowered mobile AIGC is workable in the content creation, it still faces some challenges for future popularization and development. One of the main challenges is improving blockchain performance\cite{xu2023quantum}. In public blockchains, a new block is randomly broadcast in the miner network for validation, which may lead to a large overall propagation time\cite{jiang2021taming}. Prolonged propagation delay can lead to an excessive number of forks and insufficient signature collections, affecting the performance of blockchains to manage AIGC products. 
Some efforts have been conducted to optimize the blockchain performance for the mobile AIGC networks. For example, the authors in \cite{liu2023blockchain} proposed a blockchain-based framework to manage the lifecycle of edge AIGC products, which addresses the security and trustworthiness issues of AIGC products in edge networks. The authors in \cite{wen2023freshness} introduced an unmanned aerial vehicles-enabled AIGC and formulated an incentive mechanism under information asymmetry for mobile AIGC networks, which enhances the efficiency of mobile AIGC networks. However, little work has considered how to enhance the efficiency of information propagation for blockchain-empowered mobile AIGC networks.

To address the above challenges, in this paper, we design a Graph Attention Network (GAT)-based information propagation optimization framework for blockchain-empowered mobile AIGC,
which can ultimately derive the optimal information propagation trajectory. The contributions of this paper are summarized as follows:
\begin{itemize}
    \item We introduce the Age of Information (AoI) as a metric for assessing data freshness, serving as a comprehensive measure of information propagation efficiency in public blockchains. Moreover, the AoI takes into account both information waiting and information propagation process.
    \item In pursuit of optimizing information propagation, we propose a model based on Graph Attention Networks (GAT) for information propagation optimization, in which we design a novel GAT-based architecture to minimize the overall AoI of information propagation, thereby obtaining the optimal information propagation trajectory. 
\end{itemize}

\section{GAT-based Information Propagation Optimization Framework for Mobile AIGC Networks} \label{III}

\begin{figure}[t]
    \begin{center}
		\centering
        \centerline{\includegraphics[width=0.4\textwidth]{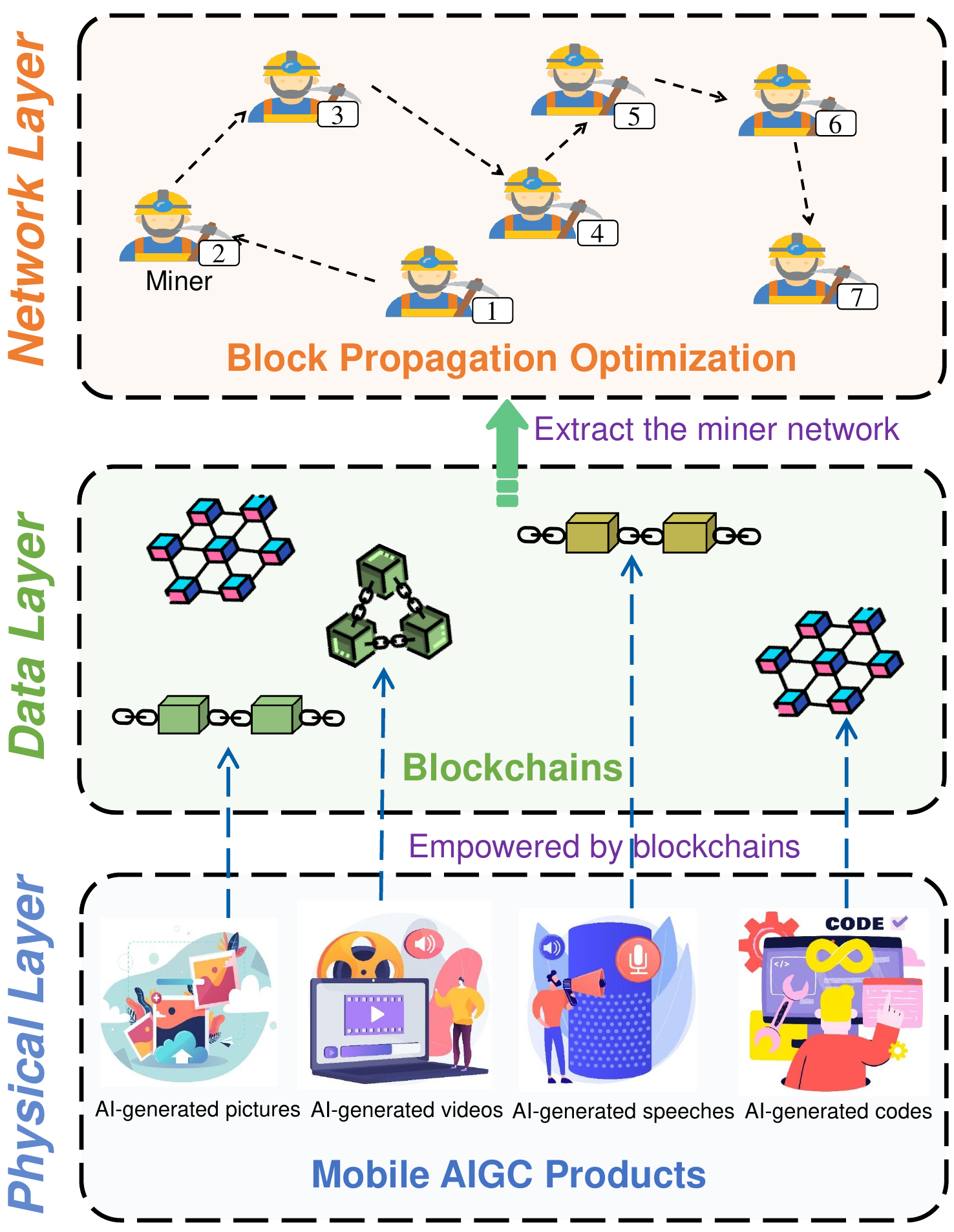}}
            \caption{A GAT-based information propagation optimization framework for blockchain-empowered mobile AIGC.}\label{framework}
    \end{center}
\end{figure}

In this section, we first briefly introduce the information propagation mechanism in public blockchains\cite{Adecker2013information}. Then, we provide a comprehensive description of the proposed GAT-based information propagation optimization framework for blockchain-empowered mobile AIGC, as illustrated in Fig. \ref{framework}. 

\subsection{Information Propagation Mechanism}
According to \cite{Adecker2013information}, we propose an information propagation mechanism in which the dissemination of new blocks diverges from the conventional flooding propagation approach inherent in traditional gossip protocols. In order to mitigate redundant transmissions to miners, a miner initiates an \textit{inv} message, comprising the hash value of the new block, to its adjacent miner prior to propagating the block. If the adjacent miner has never received the block, the adjacent miner will issue a \textit{getdata} message to request the new block \cite{Adecker2013information}, and the getdata message is then queued until the block has been received and the difficulty check has been done\cite{Adecker2013information}. To speed up information propagation, the adjacent miner sending the getdata message also sends an inv message to its adjacent miner. 

\subsection{Framework Design}
As shown in Fig. \ref{framework}, the proposed GAT-based information propagation optimization framework is structured into three layers, i.e., a physical layer, a data layer, and a network layer. Specifically, the first layer, functioning as the foundation of the proposed framework, is the physical layer, which exists in various AIGC products in mobile AIGC networks\cite{wen2023freshness,liu2023blockchain}. The second layer is the data layer, where blockchains provide secure data management and product trading circulation for mobile AIGC products\cite{liu2023blockchain}. The final layer is the network layer, which is exacted from the blockchains of the data layer. The network utilizes the GAT-based information propagation model to minimize the overall AoI of information propagation, thereby obtaining the optimal propagation trajectory. 

\section{Problem Formulation}\label{IV}
In this section, we utilize the AoI to quantify information propagation efficiency in public blockchains. The information propagation process is modeled as an M/M/1 system, operating under a first-come-first-served policy. In this paper, we consider a set $\mathcal{M} = \left\{1,\ldots, i,\ldots,j,\ldots, M\right\}$ of $M$ miners, characterized by random mobility within the miner network.

\subsection{Age of Information}

AoI is typically defined as the time elapsed since the generation of the latest successfully received message\cite{Adecker2013information},  which has been a well-established metric to capture data freshness\cite{Akosta2017age}. In this paper, we define the AoI as the time elapsed from a getdata message sent by a miner to its receipt of the new block, consisting of procedures of information waiting and information propagation\cite{wen2023task}. In particular, a lower AoI means the reduction of the time of information
propagation process, leads to a higher information propagation efficiency. 

We define the $i$-th time interval, denoted by $X_i$, as the time elapsed from miner $(i-1)$ sending the getdata message to miner $i$ sending the getdata message, which is given by
\begin{equation} \label{X_i}
    X_i=t_i-t_{i-1}.
\end{equation}
We consider $X_i$ as an exponentially distributed random variable\cite{Akosta2017age}, i.e., $X_i\sim Exp(\mu)$, where $\mathbb{E}[X_i]=1/\mu$ and $\mu$ represents the average getdata delivery ratio of miners. Then, we denote the system time of miner $i$ as $D_i$, equaling the elapsed time from miner $i$ sending a getdata message at time $t_i$ to successfully receive the block at $t_i^{'}$, which is given by
\begin{equation} \label{D_i}
    D_i=t_i^{'}-t_{i},
\end{equation}
where system time $D_i$ consists of information waiting time $W_i$ of miner $i$ and information propagation time $P_i$ from miner $(i-1)$ to miner $i$. 
According to \cite{Adecker2013information}, the AoI of information propagation between miner $(i-1)$ and miner $i$ in public blockchains is given by\cite{Akosta2017age}
\begin{equation}\label{AoI}
    \Delta_i = \frac{\mathbb{E}[X_iD_i]+0.5\mathbb{E}[X_i^2]}{\mathbb{E}[X_i]}.
\end{equation}

Considering the system time $D_i$ of miner $i$, we divide the system time $D_i$ into information waiting time $W_i$ and information propagation time $P_i$, which is denoted as
\begin{equation}
    D_i=W_i+P_i.
\end{equation}
Since $P_i$ is independent of time interval $X_i$, we can obtain
\begin{equation} \label{E[T][Y]}
	\mathbb{E}[D_iX_i]=\mathbb{E}[W_iX_i]+\mathbb{E}[P_i]\mathbb{E}[X_i].
\end{equation}
According to \cite{Akosta2017age}, we can express the information waiting time of miner $i$ as $W_i=(D_{i-1}-X_i)^+$. Given specific interarrival time $X_i = x$, the expected information waiting time is \cite{Akosta2017age}
\begin{equation} \label{WW}
	\mathbb{E}[W_i|X_i=x]=\mathbb{E}[(D-x)^+]=\int_x^\infty(t-x)f_D(t)\mathrm{d}t,
\end{equation}
where $f_D(t)$ is the Probability Density Function (PDF) of system time $D$. It has been demonstrated that the exponential distribution serves as a reasonable model for information propagation time\cite{rovira2019optimizing}, i.e., $P_i \sim Exp(1/T_i)$, where $T_i$ is the information propagation time between miner $(i-1)$ and miner $i$, calculated using the Shannon formula as follows \cite{wen2023freshness}:
\begin{equation} \label{t_p}
    T_i=\frac{B_{block}}{b\log_2\Big(1+\frac{\rho_sc^0d_{i-1,i}^{-\varepsilon}}{N_0W}\Big)},
\end{equation}
where $B_{block}$, $b$, $\rho_s$, ${d_{i-1, i}}$, $c^0$, $\varepsilon$, and $N_0$ represent the block size, the channel bandwidth between adjacent miners, the transmit power of miners, the distance between miner $(i-1)$ and miner $i$, the unit channel power gain, the path-loss coefficient, and the noise power density, respectively. For the system time $D_i$ of miner $i$, $f_{D_i}(t)$ is given by\cite{Akosta2017age}
\begin{equation}\label{ffT}
	f_{D_i}(t)=\bigg(\frac{1}{T_i}-\mu\bigg)e^{(\mu-\frac{1}{T_i})t},\: t\geq 0.
\end{equation}
Furthermore, using iterated expectation and the exponential $(\mu)$ PDF of $X_i$\cite{Akosta2017age}, $\mathbb{E}[W_iX_i]$ can be denoted as
\begin{equation}
\begin{aligned}
    \mathbb{E}[W_iX_i]&=\int_0^\infty x\mathbb{E}[W_i|X_i=x]f_{X_i}(x)\mathrm{d}x\\
    &=\frac\mu{({\frac{1}{T_i}})^3-({\frac{1}{T_i}})^2\mu}. 
\end{aligned}
\end{equation}
Therefore, based on (\ref{AoI}), the AoI between miner $(i-1)$ and miner $i$ can be given by
\begin{equation}\label{Delta_final}
\begin{aligned}\Delta_i={T_i}+\frac1\mu+\frac{\mu T_i^3}{1-\mu T_i}.\end{aligned}
\end{equation}

\subsection{Problem Formulation of GAT-based Information Propagation}
In this paper, we concentrate on achieving optimization in information propagation within public blockchains by determining the solution path of the optimal information propagation, denoted as $\boldsymbol{\pi} = (\pi_{1}, \ldots, \pi_{m})$, where $\pi_{t} \in \{1, \ldots, m\} \subseteq \mathcal{M}$ and $\pi_t \neq \pi_{t^{\prime}}$, $t \neq t^{\prime}$. It's important to note that $\boldsymbol{\pi}$ represents the set of miners involved in the optimal information propagation trajectory, and $\pi_1$ signifies the miner who secures the block consensus right for this particular block.

Based on the consensus mechanism of public blockchains, we define $m=\lfloor{M\times r^{ratio}}\rfloor$ as the number of miners participating in block validation. Here, $\lfloor \cdot\rfloor$ denotes the floor function and $r^{ratio}$ represents the proportion of the overall number of miners. Consequently, the problem of information propagation optimization is to minimize the overall AoI of information propagation, which can be formulated as 
\begin{equation} 
\begin{split}
    &\min_{\boldsymbol{\pi}}\:\sum_{i=2}^m{\Big(}\Delta_i =  T_i + \frac{1}{\mu}+\frac{\mu T_i^3}{1-\mu T_i}\Big).\\
\end{split}
\end{equation}
  Considering $T_i < (1/\mu)$\cite{Akosta2017age}, it is easy to prove that the AoI $\Delta_i$ will increase as information propagation time $T_i$ increases. Moreover, $T_i$ is a monotonic function concerning the distance $d_i$ between miner $(i-1)$ and miner $i$, according to (\ref{t_p}). Motivated by the above analysis, we propose a GAT-based information propagation optimization model to solve the optimization problem, described in Section \ref{V}.

\section{Graph Attention Network for Optimizing Information Propagation} \label{V}
In this section, we propose a GAT-based information propagation optimization model for blockchain-empowered mobile AIGC, which minimizes the overall AoI of information propagation between adjacent miners, thereby obtaining the optimal information propagation trajectory. 

GATs stand out as a potent tool by leveraging attention mechanisms to grasp the connections between miners, thereby augmenting the extraction of relational representations among them. The GAT architecture within the proposed framework comprises encoder and decoder components.
Specifically, the encoder serves to extract the structural attributes of the miner network, while the decoder systematically generates the optimal information propagation trajectory.

For the problem of information propagation optimization, $G$ as an input instance can be considered as a fully connected graph  $G = (\mathcal{M}, \mathcal{E})$, representing the miner network of public blockchains, where $\mathcal{E}$ is the edge set of the miner network. Moreover, we define $\boldsymbol{f}_i$ as the location feature of miner $i$.

For network parameters $\boldsymbol{\theta}$ and the input instance $G$, the corresponding solution probability $p_{\boldsymbol{\theta}}\left(\boldsymbol{\pi}|G\right)$ is given by
\begin{equation} \label{1}
    p_{\boldsymbol{\theta}}\left({\boldsymbol{\pi}}|G\right)=\prod_{t=1}^{m}{p_{\boldsymbol{\theta}}\left({\pi}_t |G, {{\boldsymbol{\pi}}_{1:t-1}}\right)}.
\end{equation}
Therefore, the overall route length of information propagation is given by\cite{Gkool2018GNNyuanxing}
\begin{equation}
    L(\boldsymbol{\pi}|G)=\sum_{t=1}^{m-1}\left|\left| \boldsymbol f_{\pi_t}-\boldsymbol f_{\pi_{t+1}}\right|\right|_2,\label{L}
\end{equation}
where $\left||\cdot|\right|_2$ denotes the $L2$ norm  to calculate the $2D$ Euclidean distance. The function $L(\boldsymbol{\pi}|G)$ will be utilized to train the network in Section \ref{VI}. Next, we introduce the encoder-decoder architecture constructed for the GAT-based information propagation optimization model.

\subsection{Encoder Architecture}
Firstly, the miner features are fed into a $d_h$-dimensional liner layer, where $d_h=128$. 
The initial $d_h$-dimensional miner embeddings $\boldsymbol h_{i}^{(0)}$ is calculated by\cite{Gkool2018GNNyuanxing}
\begin{equation}
    \boldsymbol h_i^{(0)}=\boldsymbol{W}^h{\boldsymbol f_i}+\boldsymbol{b}^h,
\end{equation}
where $\boldsymbol{W}^h \in \Bbb{R}^{d_h\times d_x}$ and $\boldsymbol{b}^h \in \Bbb{R}^{d_h}$ are learnable parameters.
After obtaining the embedding $\boldsymbol h_{i}^{(0)}$, it is sent to the graph attention layer and updated with $N$ GAT layers. We denote $\boldsymbol h_{i}^{(r)}$ as the miner embeddings calculated by GAT layer $r\in \{0, 1,\ldots,N\}$. Specifically, $\bar{\boldsymbol{{h}}}^{(graph)}$ is denoted as the graph embedding, which is the aggregated embedding of the input graph as the average of final miner embeddings, given by\cite{Gkool2018GNNyuanxing}
\begin{equation} \label{31}
    \bar{\boldsymbol{h}}^{(graph)}=\frac1M\sum_{i=1}^M \boldsymbol h_i^{(N)}.
\end{equation}
Finally, the encoder outputs the final miner embeddings $\boldsymbol h_i^{(N)}$, as well as the graph embedding $\bar{\boldsymbol{h}}^{(graph)}$ to the decoder.

Then, we introduce the major architecture of the proposed graph attention layer in detail. In the Multi-Head Attention (MHA) layer, the value of a miner is the compatibility of the query and the key from its neighbor\cite{vaswani2017attention}. We denote the number of attention heads as $y\in\{1,\ldots, Y\}$ and consider a sequence of query $\boldsymbol{Q}=\{\boldsymbol{q}_{1}^{(r)},\ldots,\boldsymbol{q}_{i}^{(r)},\ldots,\boldsymbol{q}_{M}^{(r)}\}$, key $\boldsymbol{K}=\{\boldsymbol{k}_{1}^{(r)},\ldots,\boldsymbol{k}_{i}^{(r)},\ldots,\boldsymbol{k}_{M}^{(r)}\}$, and value $\boldsymbol{V}=\{\boldsymbol{v}_{1}^{(r)},\ldots,\boldsymbol{v}_{i}^{(r)},\ldots,\boldsymbol{v}_{M}^{(r)}\}$. For miner $i$, $\boldsymbol q_{i}^{(r)}$, $\boldsymbol k_{i}^{(r)}$, and $\boldsymbol v_{i}^{(r)}$ are calculated by projecting the embedding $\boldsymbol h_{i}^{(r-1)}$, which are given by
\begin{equation} \label{14}
    \boldsymbol q_{i}^{(r)}=\boldsymbol{W}^Q \boldsymbol h_{i}^{(r-1)},
\end{equation}
\begin{equation} \label{33}
    \boldsymbol k_{i}^{(r)}=\boldsymbol{W}^K\boldsymbol h_{i}^{(r-1)},
\end{equation}
\begin{equation} \label{34}
    \boldsymbol v_{i}^{(r)}=\boldsymbol{W}^V\boldsymbol h_{i}^{(r-1)},
\end{equation}
where each attention head $y$ obtains parameters $\boldsymbol{W}^{Q}\in \Bbb{R}^{d_{k}\times d_{h}}$, $\boldsymbol{W}^{K}\in \Bbb{R}^{d_{k}\times d_{h}}$, and $\boldsymbol{W}^{V}\in \Bbb{R}^{d_{v}\times d_{h}}$.

Based on (\ref{14}), (\ref{33}), and (\ref{34}), we compute the compatibility $\boldsymbol u_{ij}^{(r)}\in \Bbb{R}$ by combining the query $\boldsymbol q_{i}^{(r)}$ with the key $\boldsymbol k_{j}^{(r)}$ of miner $j$ as the dot-product function\cite{vaswani2017attention}
\begin{equation}
    \boldsymbol u_{ij}^{(r)}=\begin{cases}\frac{\left(\boldsymbol q_{i}^{(r)}\right)^T\boldsymbol k_{j}^{(r)}}{\sqrt{d_k}}&\text{if miner $i$ is adjacent to miner $j$,}\\-\infty&\text{otherwise,}\end{cases}\label{23}
\end{equation}
 where the compatibility of non-adjacent miners is considered as $-\infty$. Then, based on the compatibility $\boldsymbol u_{ij}^{(r)}$, the attention weight $\boldsymbol a_{ij}^{(r)}\in[0,1]$ is given by\cite{Gkool2018GNNyuanxing} 
 
\begin{equation} \label{16}
    \boldsymbol a_{ij}^{(r)}=softmax\left(\boldsymbol u_{ij}^{(r)}\right)=\frac{e^{\boldsymbol u_{ij}^{(r)}}}{\sum_{j=1}^Je^{\boldsymbol u_{ij}^{(r)}}},
\end{equation}
where $\{1,\ldots,j,\ldots,J\}\subset \mathcal{M}$ is the adjacent miner set of miner $i$. Then, based on (\ref{34}) and (\ref{16}), the result vector $\boldsymbol h_{i}^{\prime}(r)$, combining $\boldsymbol a_{ij}^{(r)}$ with $\boldsymbol v_{j}^{(r)}$, is expressed as\cite{Gkool2018GNNyuanxing} 
\begin{equation} \label{17}
    \boldsymbol h_{i}^{\prime}{}^{(r)}=\sum_{j=1}^J \boldsymbol a_{ij}^{(r)}\boldsymbol v_{j}^{(r)}.
\end{equation}
Specifically, we use $Y=8$ heads and  $d_k=d_v=d_h/Y=16$. Besides, we define $\boldsymbol{W}_y^{O^{(r)}}\in \Bbb{R}^{d_h\times d_v}$, and the final value of the MHA layer for miner $i$ in the graph attention layer $r$ is projected to a single $d_h$-dimensional vector, given by
\begin{equation}\label{18}
    \\{MHA}_i^{(r)}\left(\boldsymbol {h}_1^{(r)},\ldots,\boldsymbol h_M^{(r)}\right)=\sum_{y=1}^Y\boldsymbol{W}_y^{O^{(r)}}\boldsymbol h_{iy}^{\prime}{}^{(r)}.
\end{equation}

    In the Batch Normalization (BN) layer, we introduce learnable parameters  $\boldsymbol{w}^{3}$ and $\boldsymbol{b}^{3}$ as the $d_h$-dimensional affine parameters, and $\overline{BN}^{(r)} (\boldsymbol h_i^{(r)})$ is denoted as batch normalization without affine transformation. Besides, we use $\odot$ to represent the element-wise product. Therefore, $BN^{(r)}(\boldsymbol h_i^{(r)})$ is expressed as\cite{Gkool2018GNNyuanxing} 
\begin{equation}
    BN^{(r)}(\boldsymbol h_i^{(r)})=\boldsymbol{w}^{3}\odot\overline{BN}^{(r)}\left(\boldsymbol h_i^{(r)}\right)+\boldsymbol{b}^{3}.
\end{equation}

In the Feed-Forward (FF) layer, we use a $d_{F}$-dimensional hidden sublayer, where $d_{F}=512$, with the parameters $\boldsymbol{W}^{f2}$ and $\boldsymbol{b}^{f2}$ and a $ReLu$ activation with the parameters $\boldsymbol{W}^{f1}$ and $\boldsymbol{b}^{f1}$ to construct the FF layer:
\begin{equation}
    FF^{(r)}\left(\hat{\boldsymbol {h}}_i^{(r)}\right)=\boldsymbol{W}^{f2}\cdot ReLu\left(\boldsymbol{W}^{f1} \hat{\boldsymbol{{h}}}_i^{(r)}+\boldsymbol{b}^{f1}\right)+\boldsymbol{b}^{f2},
\end{equation}
where the input of the FF layer $\hat{\boldsymbol {h}}_i^{(r)}$ is the output of the BN layer after the MHA layer, as calculated in (\ref{hat h}).

Finally, based on the proposed three key layers, i.e., the MHA layer, the BN layer, and the FF layer, the GAT layer is given by\cite{Gkool2018GNNyuanxing}

\begin{align}
    \boldsymbol{\hat{h}}_i^{(r)} &= BN^{(r)}\left(\boldsymbol h_i^{(r-1)}+MHA_i^{(r)}\left(\boldsymbol h_1^{(r-1)},\ldots,\boldsymbol h_M^{(r-1)}\right)\right),\label{hat h} \\
    \boldsymbol{h}_i^{(r)} &= BN^{(r)}\left(\boldsymbol{\hat  h}_i^{(r)}+FF^{(r)}\left(\boldsymbol{\hat h}_i^{(r)}\right)\right),\label{42}
\end{align}
where the layers are connected by a skip-connection\cite{Gkool2018GNNyuanxing}. Then, the result of (\ref{42}) in layer $N$ will be input to (\ref{31}) to generate the aggregated $\bar{\boldsymbol{h}}^{(graph)}$.

\subsection{Decoder Architecture}
 The decoding step for information propagation follows a sequential process. For each decoder step $t\in\{1,\ldots,m\}$, the current miner that would propagate the block selects its adjacent miner to append to the end of the sequential solution. We construct a special context vector $\boldsymbol{h}^{(d)}$ to represent the decoding context\cite{Gkool2018GNNyuanxing}, which is composed of the outputs of the encoder and decoder up to time $t$. According to \cite{vaswani2017attention}, $\boldsymbol{h}^{(d)}$ is given by
\begin{equation}
    \boldsymbol{h}^{(d)}=\begin{cases}\left[\bar{\boldsymbol{h}}^{(graph)},\boldsymbol{v}^1,\boldsymbol{v}^2\right]&\quad \text{if}\quad{t}=1,\\\left[\bar{\boldsymbol{h}}^{(graph)},\boldsymbol{v}^2,\boldsymbol{h}_{{\pi}_1}^{(N)}\right]&\quad \text{if}\quad{t}=2,\\\left[\bar{\boldsymbol{h}}^{(graph)},\boldsymbol{h}_{{\pi}_{t-2}}^{(N)},\boldsymbol{h}_{{\pi}_{t-1}}^{(N)}\right]&\quad\text{if}\quad{t}>2. \end{cases}
\end{equation}
Here, $[\cdot,\cdot,\cdot,]$ is the horizontal concatenation operator. 

In the decoder, we compute compatibility using an MHA layer similar to the encoder. After that, we use a single-head attention layer to calculate the compatibilities $u_j^{(d)}$, based on (\ref{23}). In the last step, by using the softmax, the probability of miner $i$ selected to propagate the block is given by\cite{Gkool2018GNNyuanxing}
\begin{equation}
    p_i=p_{\boldsymbol{\theta}}(\pi_t=i|G,\boldsymbol{\pi}_{1:t-1})=\frac{e^{u_j^{(d)}}}{\Sigma_{j=1}^Je^{u_j^{(d)}}}.
\end{equation}
Finally, we can obtain the solution path of the optimal information propagation trajectory based on (\ref{1}).

\subsection{Training Method}\label{VI}
In this section, we first propose an algorithm to solve the information propagation optimization problem, as shown in \textbf{Algorithm 1}. Based on (\ref{L}), the loss of the model is the expectation of the cost $L(\boldsymbol{\pi}|G)$\cite{Gkool2018GNNyuanxing}, given by
 \begin{equation}
     \mathcal{L}(\boldsymbol{\theta}|G)=\mathbb{E}_{p_{\boldsymbol{\theta}}(\boldsymbol{\pi}|G)}[L(\boldsymbol{\pi}|G)].
 \end{equation}
 We optimize $\mathcal{L}(\boldsymbol{\theta}|G)$ based on the gradient descent, which is given by\cite{Gkool2018GNNyuanxing}
 \begin{equation}
     \nabla\mathcal{L}(\boldsymbol{\theta}|G)=\mathbb{E}_{p_{\boldsymbol\theta}(\boldsymbol{\pi}|G)}\left[\left(L(\boldsymbol{\pi}|G)\right)\nabla\log p_{\boldsymbol{\theta}}(\boldsymbol{\pi}|G)\right].
 \end{equation}

\begin{algorithm}[t] \label{al}
    \caption{Gradient Descent Training-based Graph Attention Network for Information Propagation Optimization}
    \KwIn{Basic physical parameters $\small\{B_{block}, \rho_s, \varepsilon, b, N_0\small\}$, batch size $B$, number of epochs $E$, steps per epoch $T$, number of miners $M$.}
    
        Initialize $\boldsymbol\theta$.
        
    \For{$epoch = 1,\ldots,E$}
        {   \For{$step = 1,\ldots,T$}
            {
                $x_i\gets$ RandomInstance( )\cite{Gkool2018GNNyuanxing}, $\forall{i}\in\{1, \ldots, B\}$.
            
                $\boldsymbol\pi_i\gets$ Sample($G_i, \boldsymbol p_{\boldsymbol\theta}$)\cite{Gkool2018GNNyuanxing}, $\forall{i}\in\{1, \ldots, B\}$.
    
                $\nabla\mathcal{L}\leftarrow\sum_{i=1}^{B}\left(L(\boldsymbol{\pi_{i}|G)}\right)\nabla_{\boldsymbol{\theta}}\log p_{\boldsymbol{\theta}}(\boldsymbol{\pi_{i}})$.
                
                $\boldsymbol\theta \gets$ Adam($\boldsymbol\theta, \nabla\mathcal{L}$).
            }
        }
    Calculate the minimized overall AoI.
    
\KwOut {The minimized overall AoI and the optimal information propagation trajectory.}
\end{algorithm}
\begin{figure}[t]
    \begin{center}
		\centering
		\centerline{\includegraphics[width=0.4\textwidth]{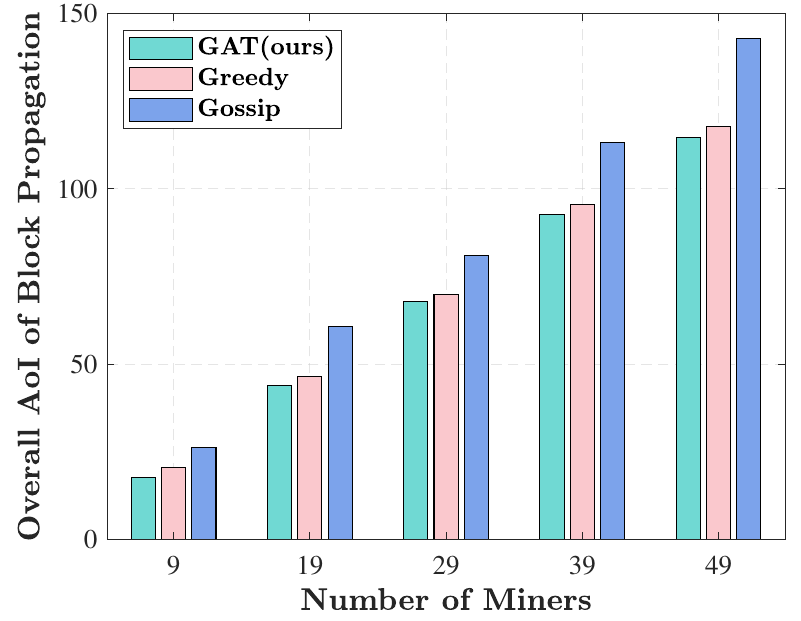}}
            \caption{The overall AoI of information propagation corresponding to different numbers of miners.}\label{AoIendd}
    \end{center}
\end{figure}

\begin{figure*}[t]
\centering

\subfigure[The information propagation trajectory in $29$ miners, which obtains $\Delta =72.45$.]
{
    \begin{minipage}[t]{0.3\linewidth}
	\centering
	\centerline{\includegraphics[width=1.2\textwidth]{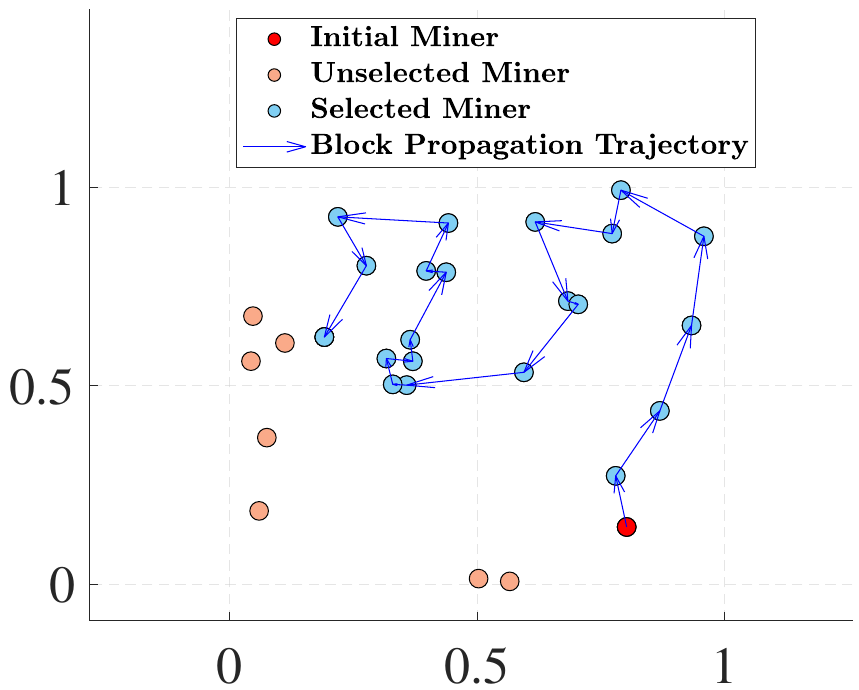}}
        \captionsetup{font=footnotesize}
	\label{29}
    \end{minipage}
}
\subfigure[The information propagation trajectory in $39$ miners, which obtains $\Delta =96.04$.]
{
    \begin{minipage}[t]{0.3\linewidth}
	\centering
	\centerline{\includegraphics[width=1.2\textwidth]{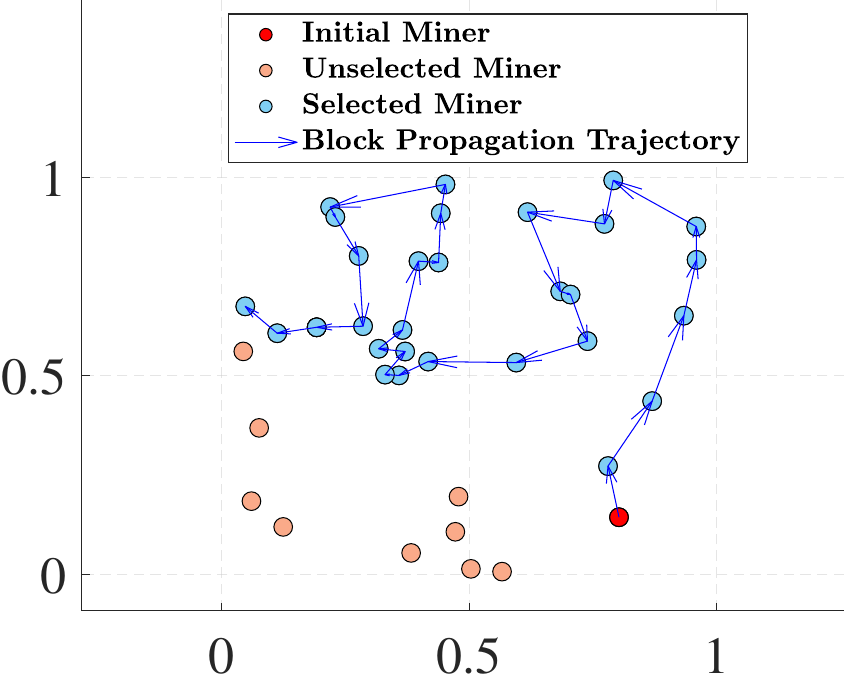}}
        \captionsetup{font=footnotesize}
	\label{39}
    \end{minipage}
}
\subfigure[The information propagation trajectory in $49$ miners, which obtains $\Delta =123.22$.]
{
    \begin{minipage}[t]{0.3\linewidth}
	\centering
	\centerline{\includegraphics[width=1.2\textwidth]{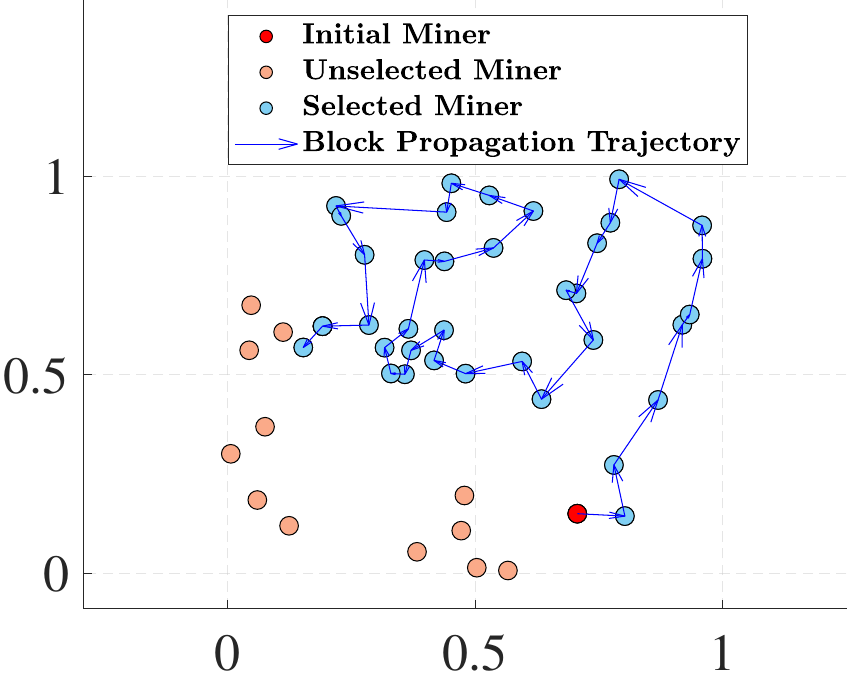}}
        \captionsetup{font=footnotesize}
	\label{49}
    \end{minipage}
}
\caption{The information propagation trajectories based on optimal AoI.}\label{trajectory}
\end{figure*}

 Note that the computational complexity of \textbf{Algorithm 1} is $O(M)$\cite{mdrori2020learning}, indicating that \textbf{Algorithm 1} is efficient. The network minimizes the overall AoI of information propagation, generating the optimal information propagation trajectory for a specific number of miners as the final output.

\section{Numerical Results}\label{VII}

In this section, we first compare the proposed GAT model with other conventional routing mechanisms: i) \textit{Greedy mechanism}, where miners select the nearest adjacent miner to propagate the block; ii) \textit{Gossip mechanism}, where miners randomly propagate the block to their adjacent miners. 

The parameters $\boldsymbol{\theta}$ of the GAT network are initialized within the range $\big(-1/\sqrt{d_f}, 1/\sqrt{d_f}\big)$, where $d_f=2$. Moreover, we use $N = 3$ layers in the encoder, which is a balance between the quality of results and computational complexity. We set $r^{ratio}=3/4$ to ensure that the information propagates to miners over $50\%$ for validation. Note that we run the experiments on NVIDIA GeForce RTX 3080Ti and we consider that the block size $B_{block}$, the channel bandwidth between adjacent miners $b$, the noise power density $N_0$, and the path-loss coefficient $\varepsilon$ are set to $1MB$, $180kHz$, $-174dBm/Hz$, and $3.38$, respectively\cite{wen2023task}.

Figure \ref{AoIendd} shows the overall AoI of information propagation corresponding to different numbers of miners utilizing different routing mechanisms. We can see that the proposed GAT model outperforms the other two mechanisms. When $M=49$, the advantage is particularly pronounced, where the AoI of the GAT model is $3\%$ lower than that of the Greedy mechanism and $26\%$ lower than that of the Gossip mechanism. The reason is that the GAT model focuses on minimizing the overall AoI of information propagation by considering the global structure of the miner network, which can obtain the optimal information propagation trajectory. In contrast, the Greedy mechanism only focuses on the selection of the current step and ignores the global structure, while the Gossip mechanism randomly selects miners without optimizing AoI, resulting in a much lower effect than other mechanisms.

In Fig. \ref{trajectory}, we present the optimal information propagation trajectory corresponding to different miner numbers $M=\{29,39,49\}$. For better distinction, we draw blue points to denote the selected miners and orange points to denote the unselected miners. Moreover, the blue arrows point to the optimal information propagation trajectories generated by the proposed GAT model. We can observe that the proposed GAT model can obtain optimal information propagation trajectories corresponding to different numbers of miners, which are clearly organized and have no unreasonable costs in AoI.

\section{Conclusion}\label{VIII}
In this paper, we have focused on enhancing the performance of blockchain-empowered mobile AIGC networks, with a particular emphasis on optimizing information propagation in public blockchains. Specifically, we have introduced the AoI as a data-freshness metric to evaluate the efficiency of information propagation. To achieve information propagation optimization, we have proposed a GAT-based information propagation optimization model to minimize the overall AoI of information propagation, thus obtaining the optimal information propagation trajectory. Numerical results have demonstrated that compared with conventional routing mechanisms, the proposed scheme can minimize the overall AoI, contributing to enhanced efficiency of information propagation in public blockchains. For future work,  we will further explore advanced variants of GATs or deeply investigate the synergy between diffusion models and GATs to better capture the interaction between miners in complex environments. 
\bibliographystyle{IEEEtran}
\bibliography{ref}
\end{CJK}
\end{document}